\documentclass[printer]{aa}
\usepackage{txfonts}
\usepackage{longtable}
\usepackage{graphicx}
\usepackage{amssymb}
\usepackage{pdflscape}

\def\aap{{Astronomy \& Astrophysics}}
\def\apj{{Astrophysical Journal}}
\def\apjl{{Astrophysical Journal Letters}}
\def\apjs{{Astrophysical Journal Supplement}}
\def\prd{{Phys. Rev. D}}

\titlerunning{Quark-Nova remnants: application to SGRs, AXPs and XDINs}
\authorrunning{Ouyed et al.}

\begin{document}

\title{Quark-Nova Remnants I:\\
 {\small The leftover debris with applications to SGRs, AXPs, and XDINs}}

\author{Rachid Ouyed, Denis Leahy, and Brian Niebergal }

\institute{Department of Physics and Astronomy, University of Calgary, 
2500 University Drive NW, Calgary, Alberta, T2N 1N4 Canada}

\offprints{ouyed@phas.ucalgary.ca}

\date{recieved/accepted}

\abstract{
We explore the formation and evolution of debris ejected around quark stars in the Quark Nova scenario,
and the application to Soft Gamma-ray Repeaters (SGRs) and Anomolous X-ray Pulsars (AXPs).  
If an isolated neutron star explodes as a Quark Nova, an iron-rich shell of degenerate matter 
forms from its crust. 
This model can account for many of the observed features of SGRs
and AXPs such as: (i) the two types of bursts (giant and regular);
(ii) the spin-up and spin-down episodes during and following the bursts
with associated increases in $\dot{P}$; (iii) the energetics of the boxing
day burst, SGR1806$+$20; (iv) the presence of an iron line as observed in SGR1900$+$14;
(v) the correlation between the far-infrared and the X-ray fluxes during the bursting episode
and the quiescent phase; (vi) the hard X-ray component observed in SGRs
during the giant bursts, and (vii) the discrepancy between the ages of SGRs/AXPs and 
their supernova remnants.
We also find a natural evolutionary relationship between SGRs and AXPs in our model 
which predicts that the youngest SGRs/AXPs are the most likely to exhibit strong bursting.  
Many features of X-ray Dim Isolated Neutron stars (XDINs) are also accounted for in our model
such as, (i) the two-component blackbody spectra; (ii) the absorption lines around
300 eV; and (iii) the excess optical emission.
 }

\maketitle

\section{Introduction}

Soft $\gamma$-ray Repeaters (SGRs) are sources of recurrent, 
short ($t \sim 0.1\,\mathrm{s}$), intense ($L \sim 10^{44}~\rm{ergs}$) 
bursts of $\gamma$-ray emission with a soft energy spectrum.
The normal patterns of SGRs are intense activity periods 
which can last weeks or months, separated by quiescent phases 
lasting years or decades.  
The three most intense SGR bursts ever recorded 
were the 5 March 1979 giant flare of SGR 0526-66 
(Mazets et al.\ 1979), the similar 28 August 1998 giant 
flare of SGR 1900+14 and the 27 December 2004 burst (SGR 1806-20).  
AXPs are similar in nature but with a somewhat weaker intensities and
no recurrent bursting.
Several SGRs/AXPs have been found to be X-ray pulsars with  
unusually high spin-down rates of $\dot{P} / P \sim 10^{-10}$~s$^{-1}$, 
usually attributed to magnetic braking caused by a super-strong 
magnetic field. 

Occasionally SGRs enter into active episodes producing many short X-ray bursts; 
extremely rarely (about once per 50 years per source), SGRs emit  giant flares, 
 events with total energies at least 1000 times higher than their typical bursts. 
Current theory explains this energy release as the result of a catastrophic 
reconfiguration of a magnetar's magnetic field. 
Magnetars are neutron stars whose X-ray emission are powered by ultrastrong magnetic fields, 
$B \sim 10^{15}$ G.  Although the magnetar model has had successes, in this paper 
we present an alternative model which addresses outstanding questions.

We explore these issues within the quark-nova (QN) scenario 
(Ouyed et al. 2002;  Ouyed et al. 2004;  Keranen, Ouyed, \& Jaikumar 2005).
 In our previous studies we have suggested that CFL (color-flavor 
locked) quark stars could be  responsible for the activity of SGRs and AXPs 
(Ouyed et al. 2005; Niebergal et al. 2006), and in this paper we extend 
the QN model by studying in more details the evolution of the QN ejecta 
 (as first discussed in Ker\"anen \& Ouyed 2003). 

Following the QN explosion we show
that a high metalicity shell can form from the neutron star crust\footnote{The 
words ``suspended crust" has been used in a different context
already with respect to quark stars; i.e. the electro-magnetically suspended crust
 at hundreds of Fermi's above the quark star surface (e.g. Alcock et al. 1986) 
instead of a few quark star radii as in our case.}.
If this matter gains sufficient angular momentum from the central quark
star it can form a torus-like structure via the propeller mechanism, which is discussed
in the second paper in this series (Ouyed et al. 2006; hereafter refered to as Paper II).
However if it does not, as we discuss in this paper, there is instead a 
thin corotating shell suspended by the quark star's magnetic field pressure.
Because the quark star is in a superconducting state it's magnetic field
decay is coupled to its period in a manner prescribed by Niebergal et al. (2006),
such that as the field decays the shell will drift closer to the star.
This movement will shift sections of the shell above the line of neutrality,
loosely defined as the polar angle (measured from the equator) above which 
the magnetic field vector is sufficiently parallel to the gravity vector to allow
material to break off from the shell, and fall into the star.  Upon collision with
the star the pieces of the shell are instantly converted to CFL quark matter 
(described by Lugones \& Horvath (2002)), 
where excess energy is released as high-energy radiation (Ouyed et al. 2004).  
We propose this as the mechanism responsible for SGR and AXP bursting activity, 
and also show that SGRs and AXPs differ primarily by age.  Another
class of objects are also explored, namely X-ray Dim Isolated Neutron stars (XDINs; e.g.~Haberl 2004),
and we find that these may have evolved from SGRs/AXPs.

This paper is presented as follows. In \S~\ref{sec:QN} we review
the concepts of a quark-nova and the expansion/evolution of the ejected material into a shell. 
\S~\ref{sec:caseA} contains discussions on the geometry of the resulting shell
and its self-similar behaviour in time, leading to pieces of the shell breaking off.
The subsequent high-energy bursting from these pieces falling into the quark star 
is then described in \S~\ref{sec:bursting}, along with methods for estimating the age of SGRs/AXPs.
In \S~\ref{sec:quiescent} the quiescent phase of SGRs/AXPs is discussed in the
context of our model, where a simple relation between luminosity and spin-down rate is derived.
The changes in period and period derivative observed during bursts are then
discussed within the framework of our model in \S~\ref{sec:periodchanges}.
\S~\ref{sec:applications} contains case studies for specific SGRs (1806$-$20 and 1900$+$14),
along with explanations for the presence of an iron line during bursts, 
correlated X-ray to infrared flux ratio, and the hard spectrum seen in giant flares.
We then summarize in \S~\ref{sec:summary} some outstanding issues in the current understanding
of SGRs/AXPs, and show how our model can provide explanations.
Finally in \S~\ref{sec:xdins}, we discuss XDINs and show how many features are readily explained
using our model.  We then conclude in \S~\ref{sec:conclusion}.

\section{Quark Nova}\label{sec:QN}

In the quark-nova (QN) scenario, the core of a neutron star (NS) shrinks to a
stable, compact, quark object before the conversion  of the entire star to (u,d,s) matter.
By contracting, and physically separating from the overlaying
material (hadronic envelope which is mostly made of crust material), the 
core drives the collapse (free-fall) of the overlaying matter leading to
both gravitational and phase transition energy release as high as
$10^{53}$~ergs in the form of neutrinos. The result is a quark star in the 
Color-Flavor Locked (CFL) superconducting phase, surrounded by an ejected shell.
Although it has been shown that pure CFL matter is rigorously electrically neutral
(Rajagopal \& Wilczek 2001), other work (Usov 2004 and references therein) indicates
that a thin crust is allowed around a quark star due to surface depletion of strange quarks.
In our model we have assumed no depletion of strange quarks, which implies a bare quark star.

In the CFL phase there are a total of nine combinations between
 the gluons and the photon with eight of these combinations subject to the Meissner effect.
 The only one that does not suffer the Meissner effect involves 
  a combination of electromagnetism and a U(1) subgroup 
of the color interactions  (e.g.  Ferrer et al. 2005). 
In other words the magnetic field will
 penetrate some phases of the  color field  but not others. 
  Unfortunately,  the relevant  calculations are done in effective models
of QCD which cannot accurately handle the details of the mixing
 and the corresponding back reaction of the quarks. 
 Assumptions had to be made (e.g. Meissner effect)
  in order to proceed  with our astrophysical model which
   implies that we are ignoring the one component that penetrates the superconductor.
In the CFL quark star, given our basic assumptions, the magnetic field is contained in the vortex array
so the internal field is uniform and equal to the surface field.

The QN ejecta consists mainly of the neutron star's metal-rich  
outer layers heated by neutrino bursts. It was shown that up to $10^{-2} M_{\odot}$ can
be released during the QN (Ker\"anen, Ouyed, \& Jaikumar  2005).
Thus our model proposes a high density, metal-rich ejecta surrounding 
a superconducting quark star.

 Within our model there are four possible scenarios as to the outcome of the ejecta.
First, if the ejecta is very light then it will become gravitationally unbound from the
quark star, where the r-process begins creating heavier elements (Jaikumar et al. 2006).
Second, if the ejecta is too heavy it will fall back into the quark star releasing tremendous
amounts of energy. This we propose could lead to explosive events
reminiscent of short gamma-ray bursts, which we are currently exploring.
Third, is the case where the ejecta mass is such that it can be suspended 
above the surface by the quark star's magnetic pressure.  Fourth, the ejecta was
formed with enough angular momentum to move into a Keplerian orbit.
The third case is discussed in this paper, whereas the fourth is discussed in the second
paper of this series.

\subsection{Expansion of the ejecta}

We consider the dynamics of the shell (of mass 
$m_{\rm sh}$) ejected during the QN.
If the shell's outward momentum is not too large, the interplay
between gravity and the QS's magnetic field is dominant.
Thus, as the highly conducting shell expands spherically
outwards, an electromotive ($\vec{J}\times\vec{B}$) force is induced 
to oppose its motion, causing it to continuously decelerate\footnote{
The damping term is by far the dominant term in the equations of motion
of the shell due to the high conductivity of the shell, estimated by 
$\sigma = n_{\rm e,th}e^2\lambda/\left(m_{\rm e}c_s\right)$. 
Here $n_{\rm e,th}$ is the number density of thermal electrons,
$c_s$ is the sound speed, the mean free path is given by
$\lambda = 1/\left(n_{\rm e,th}\sigma_{\rm T}\right)$,
and $\sigma_{\rm T}$ is the Thompson scattering cross section.}.
The natural oscillatory motion created by the magnetic 
pressure gradient and gravity is resultingly damped out,
resulting in the ejecta finding an equilibrium radius
where the forces due to the magnetic pressure gradient and gravity
are in balance.  This equilibrium radius is then given by,
\begin{equation}
R_{\rm m} \simeq 10\   {\rm km}\sqrt{\sin\theta_{\rm B}}\  \frac{ B_{\rm s, 15}R_{\rm QS, 10}^3}
                {\sqrt{m_{\rm sh, -6} M_{\rm QS, 1.4}}}   \ .
\label{aradius}        
\end{equation}
Here we have used $B_{\rm m} = B_{\rm s} (R_{\rm s}/R_{\rm m})^3$ for a dipole 
magnetic field with a strength at the surface
of, $B_{\rm s, 15}$, in units of $10^{15}~\rm{G}$. 
The star's mass and radius are in units of $1.4 M_{\odot}$ and $10~\rm{km}$ respectively,
while the mass of the shell, $m_{\rm sh, -6}$, is in units of $10^{-6} M_{\odot}$.  
The line of neutrality, $\theta_{\rm B}$, is measured from the equator and
defines the angle above which the gravity vector is no longer mostly perpendicular 
to the magnetic field vector.  If sections of the shell are above this line, they are
free to break off and fall into the star's poles along the field lines.
The condition for the shell to exist is that it is off the surface of the quark star, 
$R_{\rm m} > R_{\rm QS}$, which implies 
$B_{\rm s, 15}^2 R_{\rm QS, 10}^4 \sin\theta_{\rm B} > (m_{\rm sh, -6} M_{\rm QS,1.4})$.
 
For this equilibrium situation to be achieved 
the mass and initial velocity of the shell are
confined to a somewhat narrow range of values.
Although the mass of the shell is self-consistent with
previous studies of neutron star crust masses (discussed
below), the initial velocity is not similarly constrained.  
As such, we assume it can take on a range
of values, with the small initial velocity case 
corresponding to the work presented in this paper,
and the high velocity case is left for the second
paper in this series.  We feel that this dichotomy is 
physical, and explains some observed AXP features well.
 
 Recent work studying natural mechanism of magnetic
field amplification in quark matter (just before or during the 
onset of superconductivity; Iwazaki 2005), shows 
 that $10^{15}$ G  magnetic fields are readily achievable.
  This amplification 
can occur in quark matter due to the response of quarks to the spontaneous magnetization of the gluons. 
 In contrast, for the magnetar model, the sole proposed mechanism 
for generating $10^{15}$ G fields requires millisecond proto-neutron stars.
 This seems to be challenged by recent observations (Allen \& Hovarth 2006).

Thus, equation~(\ref{aradius}) implies that for magnetic fields in the range of $10^{14}$-$10^{15}~\rm{G}$,
the maximum mass of the shell cannot exceed $10^{-8}$-$10^{-6}~M_{\odot}$.
A higher shell mass translates into $R_{\rm m} < R_{\rm QS}$, meaning that the magnetic field is
not sufficiently strong to stop the shell from promptly falling back onto the star.  
This situation is very possible but not considered in this paper, as it is left for future work.

The shell's size  is determined by the NS crust density profile (Datta et al. 1995),
and is of the order of ($0.012$-$0.025)]\times R_{\rm QS}$ corresponding to shell masses
of the order of $(10^{-8}$-$10^{-6})M_{\odot}$ for densities of
$(1.8\times 10^{8}$-$5.5\times 10^{10})$ g/cc.  We note that if degenerate, 
the shell will always be in relativistic degeneracy since
the densities are above $\rho_{\rm cr} = 7.3\times 10^{6} \mu_{e}$ g/cc (Lang 1974).  Here,
$\mu_{\rm e}$ is the
 mean mass per electron and is taken to be $\mu_{\rm e}\approx 2$ since the 
shell's maximum density is below the neutron drip density, $\sim 4\times 10^{11}$ g cm$^{-3}$ .

The transition from Fermi-Dirac to Boltzmann statistics occurs at the degeneracy temperature
$T_{\rm fermi} \simeq  76.2~\rm{MeV}~\rho_{\rm 10}^{2/3}$
for a non-relativistic gas and $T_{\rm fermi} \simeq  8.8~\rm{MeV}~\rho_{\rm 10}^{1/3}$  for a relativistic gas (e.g. Lang, p 253).
So, for the shell to be born degenerate at 10 MeV, it implies a minimum density
of $\sim 10^{10}$ g/cc which translates to an ejected mass of $\sim 10^{-7}M_{\odot}$.
We note that even if the shell density at birth is less than $\sim 10^{10}$ g/cc, 
(blackbody) cooling during the expansion of the ejecta to $R_{\rm m}$ results in a shell
temperature of  less than $1$ MeV, thus, yielding a degenerate shell.

Once the shell reaches the magnetic radius $R_{\rm m}$, it will have expanded  
 to a thickness of, $\Delta R_{\rm m}$, such that it satisfies hydrostatic equilibrium.  
Using the relativistic degenerate equation-of-state ($P = \varkappa\rho^{4/3}$ where 
 $\varkappa = 1.244\times 10^{15} \mu_{\rm e}^{-4/3}$), 
the width of the shell, positioned at $R_{\rm m}$, can be calculated to be,
\begin{equation}
 \frac{\Delta R_{\rm m}}{R_{\rm m}} = 2.1\times 10^{-2} 
 m_{\rm shell, -6}^{1/4} M_{\rm QS, 1.4}^{-3/4}\  .
 \label{eq:drm}
\end{equation}
 This implies that the shell's thickness does not change much when it reaches $R_{\rm m}$.
 The corresponding average density of the shell at the magnetic radius is then,
\begin{equation}\label{eqn:density}
 \rho_{\rm m} = 7.58\times 10^{9}~\rm{g/cc}\  m_{\rm sh, -6}^{9/4}  
 M_{\rm QS, 1.4}^{9/4} B_{\rm s, 15}^{3} R_{\rm QS, 10}^9 \ ,
\end{equation}
implying that the shell is relativistic degenerate.

\subsection{Energy release}

The internal energy  of the expanding ejecta is 
 $ U = N (3/5)\epsilon_{\rm F}
\left[1 + \frac{5\pi^2}{12}\left(\frac{kT}{\epsilon_{\rm F}}\right)^2\right] $
where $N= m_{\rm shell}/\mu_{\rm e} m_{\rm H}$, is the total  number of  electrons.
 The Fermi energy in this case is $\epsilon_{\rm F} \sim 8.8\ {\rm MeV} \rho_{10}^{1/3}$ (e.g. Pathria pg.200). 
As a result of the heating of the electrons, the optically thick shell will radiate 
approximately as a blackbody,
\begin{equation}
L_{\rm sh} = 4\pi R_{\rm sh}^2 \sigma T_{\rm sh}^4 \ .
\end{equation} 
The total energy it releases as it relocates to $R_{\rm m}$ is  then (taking $kT << \epsilon_{\rm F}$)
\begin{equation}
\Delta U_{\rm F} \sim 5.2\times 10^{45}\ erg \  m_{\rm sh, -6} (\rho_{i,10}^{1/3} - \rho_{m,10}^{1/3})\ ,
\label{eq:qnefrho}
\end{equation}
where $\rho_{i,10}$ and $\rho_{m,10}$ are the densities of the shell initially and at 
magnetic equilibrium radius respectively, in units of $10^{10}~\rm{g/cc}$.
This energy released during expansion is done so in roughly a millisecond.

We have not
investigated the fate of this energy release, whether it is absorbed by adiabatic
 expansion losses or radiated. 
 What is crucial to our model is the fact that the shell remains
degenerate at $R_{\rm m}$ which is guaranteed by hydrostatic equilibrium.

\subsection{Angular momentum transfer}

For a magnetic radius, $R_{\rm m}$, larger than the corotation radius the propeller 
will take effect (Schwartzman 1970; Illarionov \& Sunyaev 1975),
 deflecting the QN shell into a torus on the equatorial plane.
Using an angular momentum conservation argument, we can estimate the location
of such a torus by writing 
\begin{equation}
R_{\rm m}^2 \Omega_{\rm QS} = R_{\rm t}^2 \Omega_{\rm K}\ ,
\end{equation}
where $R_{\rm t}$ is the equatorial location of the torus in a Keplerian
orbit around the quark star, and $\Omega_{\rm K} = \sqrt{GM_{\rm QS}/R_{\rm t}^3}$ 
is the Kepler rotation.
Applying equation~(\ref{aradius}) implies a torus radius of,
\begin{equation}\label{dradius}
 R_{\rm t} = \frac{R_{\rm m}^4 \Omega_{\rm QS}^2}{G M_{\rm QS}} 
 \simeq 0.2\ {\rm km} \frac{\sin^2\theta_{\rm B} B_{\rm s, 15}^4 R_{{\rm QS,}10}^{12} }
 { m_{\rm sh, -6}^{2} M_{\rm QS, 1.4}^3 P_{{\rm i,ms}}^{2}}  \ ,
\end{equation}
where the birth period, $P_{\rm{i,ms}}$, is given in units of milliseconds.
    
However, in order to actually form a torus we require enough angular momentum 
transfer to guarantee $R_{\rm t} > R_{\rm m}$. This translates into an upper limit
on the initial period of,
\begin{equation}
P_{\rm{i,lim}} < 0.46~{\rm ms}~
\frac{\left(\sin\theta_{\rm B}\right)^{3/4} B_{\rm{s,}15}^{3/2} R_{{\rm QS,}10}^{9/2} }
{m_{\rm{sh,}-6}^{3/4} M_{\rm{QS,}1.4}^{5/4}} \ .
\label{eq:plimit}
\end{equation}

This presents two cases: we study the first
 in this first paper  when
the star's initial period is too large for torus formation.
This is where the shell will remain at $R_{\rm m}$ and be in corotation with the magnetosphere.
The second case where the shell can achieve 
a Keplerian orbit and form a torus is studied in paper II (Ouyed et al. 2006).

 \section{Co-rotating Degenerate Shell}\label{sec:caseA}

\subsection{Shell geometry}
 
The area of a shell at the magnetic equilibrium radius, $R_{\rm m}$, is 
$A_{\rm sh} = 4\pi R_{\rm m}^2 \sin\theta_{\rm B}$, 
where $\theta_{\rm B}$ is the polar angle (measured from the equator) and defines the line of neutrality. 
In other words, for $\theta > \theta_{\rm B}$ the shell material is free to 
``slip'' along the magnetic field lines onto the star, whereas for $\theta < \theta_{\rm B}$ the
shell material is suspended by magnetic pressure.
Thus, the geometry is such that there is a thin shell at the equator subtending an angle of 
$2\theta_{\rm B}$, and empty regions at the poles.  The shell is shaped in the star's distorted 
magnetic dipole with an inward bulge at the equator, and the radius of the shell at the 
equator, under hydrostatic balance, can be shown to be roughly half that at $\theta_{\rm B}$.

\subsection{Self-similarity: shell dynamics and evolution}\label{sec:self_sim}

The radius of the shell, $R_{\rm m}$, is proportional to the magnetic field strength 
(i.e.~Eq.~\ref{aradius}), so we use the magnetic field decay prescribed by 
Niebergal et al. (2006) to calculate the radius in time.  This is because the quark 
star itself is in the superconducting phase, implying that vortex expulsion couples
field decay and period evolution.

After a characteristic time, $\tau_{\rm n}$ (as defined by Eq.~4 in Niebergal et al. 2006),
the shell will respond to the decaying field by moving towards the 
star, resulting in a new equilibrium radius.
During this movement, sections of the shell will be shifted above the 
line of neutrality, $\theta_{\rm B}$, where they will be broken off
and fall into the star along the magnetic field lines.  Upon colliding with
the quark star these shell pieces will be converted immediately to CFL quark matter,
releasing excess energy in the form of radiation, which in our model is an SGR/AXP burst.
We then have a new state, $n+1$, where the shell has moved towards the star, lost mass, 
and is now sitting at radius $R_{\rm m, n+1}$ with a mass $m_{\rm sh, n+1}$.  

The new characteristic decay time, period of rotation, period derivative, 
and magnetic field are respectively,
\begin{eqnarray}
\label{eq:tau}\tau_{n+1} &=& \kappa \frac{P_{n+1}^2}{B_{n+1}^2} \\
\label{eq:p_evol}P_{n+1} &=&  P_{n} (1+\frac{t}{\tau_{n}})^{1/3} \\
\label{eq:pdot_evol}\dot{P}_{n+1} &=&  \frac{P_{n}}{3\tau_{\rm n}} (1+\frac{t}{\tau_{n}})^{-2/3} \\
\label{eq:b_evol}B_{n+1} &=& B_{n} (1+\frac{t}{\tau_{n}})^{-1/6} \ ,
\end{eqnarray}
where the constant $\kappa = 8.8\times 10^{38}~\rm{G}^2\cdot\rm{s}^{-1}$, is for a quark star with
a mass of $1.4M_{\odot}$ and radius of $10~\rm{km}$. Also, $t$, is
the time elapsed since the previous burst (or birth of the quark star if it is the case).
Taking the ratio of the new characteristic time to the old gives,
\begin{equation}\label{eq:tau_ratio}
\frac{\tau_{n+1}}{\tau_n} = 1 + \frac{t}{\tau_n} = 1+ \alpha\ ,
\end{equation}
where $\alpha \equiv t/\tau_n$ defines the typical decay time of the magnetic field
for each epoch $n$. Because of the self-similar nature of the problem, $\alpha$
is roughly constant from one epoch to another, which implies 
that the time interval between two successive bursts,
\begin{equation}
\Delta t_{n+1} = \alpha \tau_{n+1} = \alpha (1+\alpha)^{n}\tau_{n} \ ,
\end{equation}
increases in time on average.

So after a time $t= \alpha \tau_{n}$, the magnetic field decays and the shell
originally sitting at $R_{\rm{m,n}}$  moves in closer to the star.  
Now by keeping the mass of the shell constant as it slowly drifts in from $R_{\rm m}$ to an inner
radius $R_{\rm in}$ and applying equation~(\ref{aradius}) we can write,
\begin{equation}\label{eq:constmass}
R_{\rm in}/R_{\rm m} = B_{n+1}/B_{n} = (1+\alpha)^{-1/6} \ .
\end{equation}
In the above equation the notation used implies that the shell reaches
$R_{\rm in}$ when the surface magnetic field has decayed to $B_{\rm n+1}$.
Following this, the shell will then suddenly lose mass at $R_{\rm in}$, due to the pieces of the shell
closest to the poles being shifted above the line of neutrality, $\theta_{\rm B}$, 
and breaking off. The resulting  change of mass of the shell can be 
realized by considering the area of shell material below the line of neutrality
while keeping density in the shell roughly constant during the process,
\begin{equation}\label{eq:massratio}
\frac{m_{\rm{sh,}n+1}}{m_{\rm{sh,}n}} = \frac{4\pi \sin\theta_{\rm B} R_{\rm in}^2}{4\pi \sin\theta_{\rm B} R_{\rm m, n}^2} = \frac{R_{\rm in}^2}{R_{m, n}^2} = (1+\alpha)^{-1/3} \ .
\end{equation}
The shell will now be less massive causing it to rapidly 
(i.e.~keeping $B$ constant, $B_{\rm n+1}= B_{\rm in}$)
equilibriate by moving further away from the star to a new radius that defines
the $(n+1)$ state.  It so happens, this new radius is the same as the original.
 This is seen from equation~(\ref{aradius}),
\begin{equation}
 \frac{R_{\rm{m,} n+1}}{R_{\rm{m,}n}} 
 = \frac{B_{n+1}}{B_{n}} \sqrt{\frac{m_{\rm{sh,}n}}{m_{\rm{sh,}n+1}}} = 1 \ .
\end{equation}
Therefore as the shell adjusts over time, it stays at roughly the original radius
$R_{\rm m, n}$.  In other words, it oscillates about the same position while losing mass.
This mass loss can be determined appealing to 
equations~(\ref{eq:constmass} \& \ref{eq:massratio}), 
\begin{equation}\label{eq:mloss}
 \Delta m_{\rm{sh,}n} = m_{\rm sh, n} - m_{\rm sh, n+1}= f\left(\alpha\right) \times m_{\rm{ sh},n} \ ,
\end{equation}
where we define $f(\alpha)$ as,
\begin{equation}\label{eq:falpha}
f\left(\alpha\right) = \left(1-\left(1+\alpha\right)^{-1/3}\right) \ .
\end{equation}
Thus, the amount of mass lost during a burst in this self-similar model (Eq.~\ref{eq:mloss})  
is determined by $\alpha$. 
 The corresponding energy release is,
\begin{eqnarray}\label{eq:eburst}
 E_{\rm{b},n} & = & \eta \Delta m_{\rm{sh,}n} c^2  
   = \eta m_{\rm{sh,}n}c^2 f\left(\alpha\right) \\ \nonumber
   &=&  1.8\times 10^{47} \  {\rm erg}\  \eta_{0.1} m_{\rm{sh,}-6} f(\alpha) \ ,
\end{eqnarray}
where $\eta_{0.1}$ is the energy conversion efficiency factor taken to be $\sim 0.1$.
 We also note that the ratio between succesive burst energies is,
 \begin{equation}
  \label{eq:erecur}
 \frac{E_{\rm b, n+1}}{E_{\rm b, n}} = \frac{m_{\rm sh, n+1}}{m_{\rm sh, n}} = (1+\alpha)^{-1/3}\ ,
\end{equation}
which implies that the bursts weaken in time. This
is expected since the shell's mass is decreasing in time.

In our model, AXPs are merely older versions 
of SGRs, thus one would expect SGR bursts to 
be more intense and frequent. As can be seen in 
table \ref{Observations} (where the objects are in order of 
decreasing estimated age), there is seemingly 
a decreasing trend in burst intensity and 
frequency as one moves from younger to 
older objects. It may be that both types of bursts have this trend, 
however, we do not feel there is enough events to draw any conclusions.
This is a potential test for our model.

\section{The bursting phase}\label{sec:bursting}

The magnetic energy due to field decay is released continuously over
a long timescale (thousands of years) and gives the steady x-ray luminosity 
of SGRs and AXPs (see \S \ref{sec:quiescent}). What is
important and unique about the shell (i.e. the accretion energy) is that the energy is released
 in bursts, and the shell sitting
  at $R_{\rm m}$ offers a natural mechanism/torque for sudden changes
in period derivative.

After time $t\sim \alpha \tau$, the magnetic field decays substantially 
and the entire shell moves in to $R_{\rm in}/R_{\rm m}\sim (1+\alpha)^{-1/6}$.
A larger $\alpha$ means the shell moves closer to the star, 
causing larger sections to be shifted above the line of neutrality, 
implying larger bursts\footnote{If $R_{\rm in} \le R_{\rm QS}$ then 
areas of the shell are in contact with the star. 
In this case, it is easy to imagine that  
the shell will experience a major disruption as inner sections are converted
to CFL matter during contact probably destroying the entire
shell; this could have applications to other explosive
phenomena  and is beyond the self-similar picture presented here.}.

 Another possible scenario is that 
  giant bursts are due to global instabilities like Rayleigh-Taylor while
 regular bursts are due to chunks breaking-off the edge of the shell.
 We assume that the Rayleigh-Taylor instability does not act due
to the long timescale for magnetic penetration of the conducting shell (see \S  4.1
 in paper II).
  What is presented below is intended to present the overall  scenario and follow-up
   work (i.e. numerical simulations) is necessary to investigate the effects global instabilities might have 
    on this simplified picture.

\subsection{Giant bursts: $\alpha \simeq 1$}

If the time needed for the shell to move in sufficiently to have pieces broken off
is roughly equal to the charactertic field decay time, then $\alpha=1$, and $\sim20$\% 
of the shell's mass is lost during the burst. 
According to equation~\ref{eq:eburst}, the corresponding energy release is,
\begin{equation}\label{eq:egburst}
 E_{\rm{b},n}  =  3.6\times 10^{46} \  {\rm erg}\  \eta_{0.1} m_{\rm{sh,n,}-6} f_{0.2} \ ,
\end{equation}
where $f_{0.2} = f(\alpha =1)$ and $m_{\rm{sh,n,}-6}$ is the current mass of the shell in units
of $10^{-6}M_{\odot}$.  The waiting time between each bursts is on average,
\begin{equation}\label{eq:bfreq}
 \Delta t = \alpha \tau \approx \alpha\times 2500~{\rm yr}\ \left(\frac{P_{10}^2}{B_{15}^2}\right) \ ,
\end{equation}
where $P_{10}$ is the spin period in units of $10~\rm{s}$, and the magnetic field is in
units of $10^{15}~\rm{G}$.  One can see here that, because $B$ decreases and $P$ increases
in time, the waiting times are increasingly less frequent as the object ages.
Given the range in $B$ derived for SGRs in our model (see Table 1), this
implies a burst frequency of $\simeq 1/(\alpha 1000\ {\rm yr})$ per object.

\subsection{Regular bursts: $\alpha \ll 1$\label{sec:rbursts}}

If pieces of the shell are able to break off as the shell
moves in by only a small amount, then $\alpha \ll 1$.
This corresponds to small pieces of the shell falling into the star and,
in our model, leads to regular-sized bursts.
This process can be interpreted as small oscillations around $R_{\rm m}$ 
consistent with small fractional mass-loss episodes by the shell. 
Observationaly, regular bursts are known to follow a power-law distribution in energies
which in our model would be due to a power-law distribution in $\alpha$.
Clearly, the detailed mechanism which determines $\alpha$ would be different 
for giant bursts and regular bursts, and would require numerical simulations to understand.

For $\alpha \ll 1$, equation~(\ref{eq:mloss}) approximates to 
$\Delta m_{\rm{sh,}n} \simeq (\alpha/3) m_{\rm{sh,}n}$ 
which yields burst energies of,
\begin{equation}\label{eq:eregburst}
 E_{\rm b} \sim \alpha\times ( 6\times 10^{46}\ {\rm erg~s}^{-1})~ \eta_{0.1} m_{\rm s,n,-6} \ .
\end{equation}
For $\alpha = 10^{-6}$, a burst energy of $E_{\rm b}\sim 6\times 10^{40}$ erg is attained,
and one should expect a waiting time of $\Delta t \sim 1$ day (see Eq.~\ref{eq:bfreq}), 
for a star with a $7~\rm{s}$ rotation period and field strength of $10^{15}~\rm{G}$.

Thus, in our model the difference between regular and giant bursts 
is due to the size of piece that breaks off the shell. Although the size is 
of a stochastic nature, we expect younger objects (i.e. the SGRs)
to have a more unstable shell, resulting in more
large pieces breaking off, causing more frequent 
and intense bursts.  For older objects (i.e. the AXPs) the shell becomes 
more stable with age resulting in weaker, less frequent, bursts.

 \begin{figure*}[t!]
\includegraphics[width=1.\textwidth]{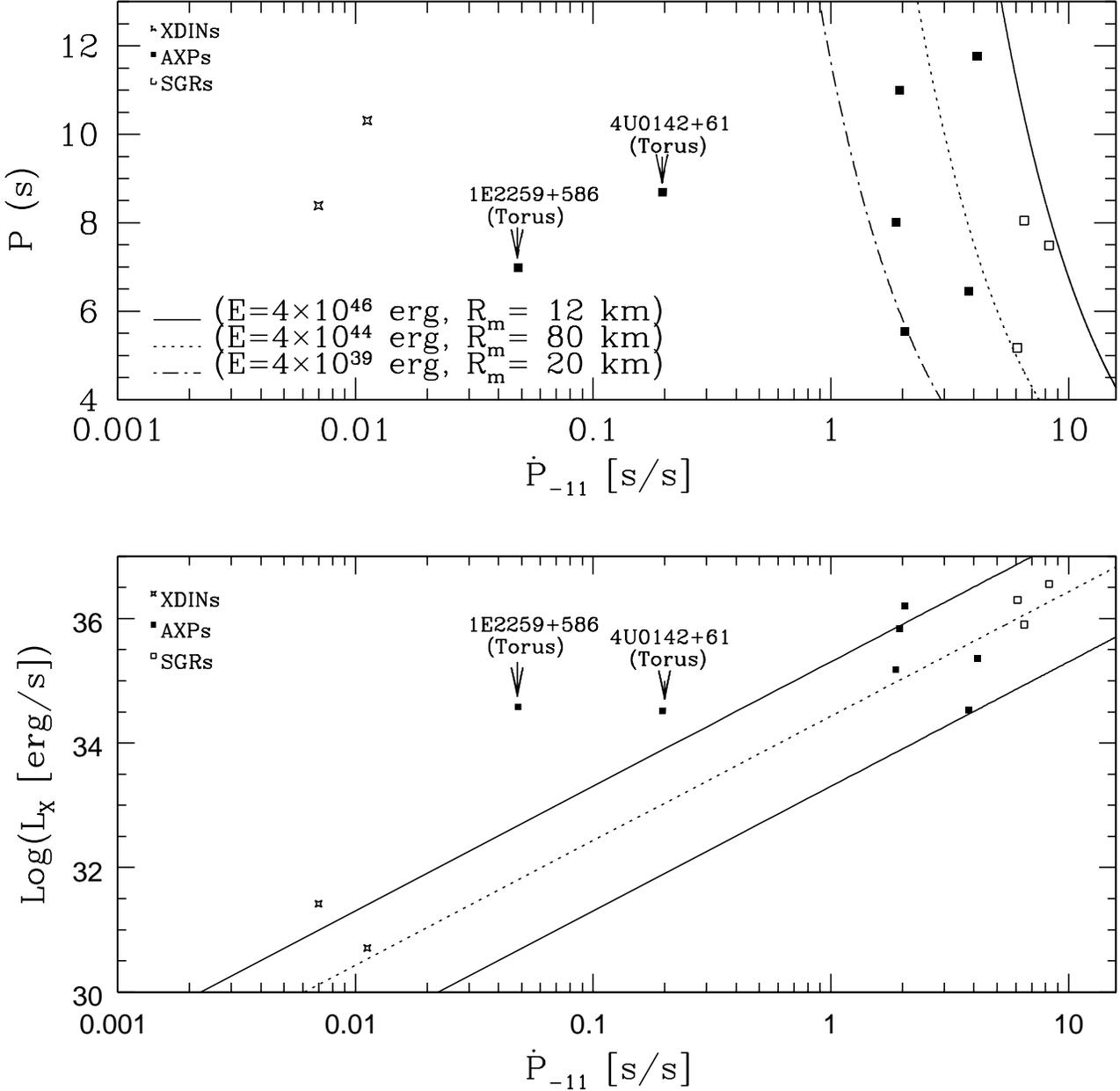}
\caption{\label{fig:ppdot}
The upper panel is a $P$-$\dot{P}$ diagram with SGRs/AXPs and XDINSs plotted.
 The solid curve is a plot of Eq.~(\ref{eq:ppdot}) with a burst energy of $E_{\rm b}= 10^{46}$ erg, 
 $\alpha = 2$, and $R_{\rm m}= 20$ km. 
Also shown are curves for $R_{\rm m}=100$~km, 
$\alpha =2$ and energies $E_{\rm b}= 10^{44}$ erg~s$^{-1}$ (dotted) and 
$E_{\rm b}= 10^{41}$ erg~s$^{-1}$ (dashed).  
These line of constant energy move leftwards as an object ages.
In the lower panel luminosity is plotted against period derivative according to Eq.~\ref{eq:lx}.
The upper and lower solid lines represent a magnetic to X-ray conversion efficiency, 
$\eta_{\rm X}$, of 1 and 0.01 respectively,
and the dashed line is the luminosity averaged over all viewing angles (for $\eta_{\rm X}=1$). 
Since $\dot{P}\propto 1/\rm{age}$, objects further to the left are generally older. }
\end{figure*}

\subsection{The $P$-$\dot{P}$ diagram}

From equations~(\ref{eq:p_evol} \& \ref{eq:b_evol}),  the magnetic
field is related to the period and period derivative by,
\begin{equation}\label{eq:3kppdot}
B = \sqrt{3 \kappa P\dot{P}} \ .
\end{equation}
Furthermore, combining this and the equation for $R_{\rm m}$ (Eq.~\ref{aradius}) with 
equation~(\ref{eq:eburst}) one can show that,
  \begin{equation}
  P \dot{P}_{-11} \simeq 21 \frac{R_{\rm m,10}^2}{\sin\theta_{\rm B}}\times  \frac{E_{\rm b, 47}}{\eta_{0.1} f\left(\alpha\right)}\ ,
  \label{eq:ppdot}
  \end{equation}
 Illustrated in  the upper panel of Figure~\ref{fig:ppdot}, are cases with large and small $\alpha$. 
 The solid curve is for $E_{\rm b}= 10^{46}$ erg~s$^{-1}$, $\alpha = 2$, and $R_{\rm m}= 20$ km,
 and fits well with the boxing day event. Also shown are two curves for $R_{\rm m}=100$ km, 
 $\alpha =2$ and energies $E_{\rm b}= 10^{44}$ erg~s$^{-1}$ (dotted) and 
$E_{\rm b}= 10^{41}$ erg~s$^{-1}$ (dashed).
The lines of constant burst energies in the upper panel of Figure \ref{fig:ppdot}
move to the left as time increases (as the object ages).  This can be seen
 from equation (\ref{eq:erecur}) in 
 \S~\ref{sec:self_sim}   which shows the burst energy decreasing in time.
 We also note that as objects evolve further to the left, the bursts
 are weaker and less frequent, which is why AXPs in our model evolve from SGRs.

\subsection{Age estimate}\label{sec:ageestimate}

To determine the exact age of AXPs/SGRs in our model, one would need to know the entire
bursting history of the object which is impossible.
For a real situation the burst sizes are variable implying a different
$\alpha$ for each burst. So to approximate the age in the simplest case 
(when $\alpha$ is constant over time on average), the age elapsed since the QN is,
\begin{equation}
t_{\rm QN} = \sum_{i=1}^{n} \alpha \tau_{\rm  i}\ .
\end{equation}
Using equation~(\ref{eq:tau_ratio}) the series can be summed to yield
\begin{equation}\label{eq:age}
t_{\rm QN} = \tau_{\rm n} \left(  1 - (1+\alpha)^{-n}\right)\ .
\end{equation}
Provided that there has been a large number of bursts ($n\gg1$), 
\begin{equation}\label{eq:est_age}
\tau_{\rm n}=P_{\rm n}/3\dot{P}_{\rm n}
\end{equation} 
is a good estimate of the age, however we do expect $\alpha$ to vary from
one burst to another. 
Although for SGRs, the energy in the large bursts ($\alpha\sim 1$) dominates 
the small bursts ($\alpha\sim 10^{-6}$), resulting in an age very close to 
that of equation~(\ref{eq:age}).
This age also happens to be the age estimate derived from 
equations~(\ref{eq:tau} - \ref{eq:pdot_evol}) when assuming $t\gg\tau$, 
which is reasonble for AXPs in our model.
So our model predicts the age of $\sim 0.5 P/3\dot{P}$ for SGRs, 
and $\sim P/3\dot{P}$ for AXPs and XDINSs.
However, without knowing the burst history we cannot derive the initial
period and magnetic field (see equations \ref{eq:p_evol} and \ref{eq:b_evol}).

One may be tempted to use the age of the associated parent supernova remnant to
estime the age of the SGR/AXP, but in our model these ages are neccesarily different.
This is due to the time required for a neutron star to reach quark deconfinement densities
(Staff et al. 2006), as well as the time needed for strange quark nucleation 
(i.e.~\textit{u,d} to \textit{u,d,s}) to occur (Bombaci et al. 2004).  
Together these delays can easily add up to the age difference between SGRs/AXPs and their
supernova remnants.
     
\section{The quiescent phase}\label{sec:quiescent}

The X-ray luminosity during the quiescent phase of SGRs/AXPs in our model
is due to vortex expulsion from spin-down.  The magnetic field contained within
the vortices is also expelled, and the subsequent magnetic reconnection leads
to the production of X-rays. From equation~(22) in Ouyed et al. (2004) and
equations~(\ref{eq:p_evol}-\ref{eq:b_evol}), the luminosty emitted from this process is,
\begin{equation}\label{eq:lx}
L_{\rm X} \simeq 2.01\times 10^{35}\ {\rm erg s}^{-1} \eta_{\rm X} \dot{P}_{-11}^2 \ ,
\end{equation} 
where $\eta_{\rm X}$ is the efficiency parameter inherent in the conversion from magnetic
energy to radiation.  The lower panel in Figure~\ref{fig:ppdot} shows our 
model of luminosty versus period derivative as compared to the quiescent X-ray 
luminosities of AXPs, SGRs and XDINSs.

The scatter in the AXPs/SGRs data can be explained within our model
by considerations of shell geometry and viewing angle.  
More specifically, any X-rays being emitted from the star's surface towards the 
solid angle filled by the shell ($4\pi\sin\theta_{\rm B}$) will be affected
due to absorption by the shell.
X-rays are then able to escape from an open area at the poles with a solid angle on the order of 
$\left(4\pi (1-\sin\theta_{\rm B})\right)$.
Because we statistically observe viewing angles equally from all directions, this would imply
an average (geometrical mean) lumimosity of $L_{\rm X}\times (1-\sin\theta_{\rm B})
\sim 2.7\times 10^{34}\eta_{\rm X}\dot{P}_{-11}\ {\rm erg~s}^{-1}$, 
assuming $\theta_{\rm B}=60^{\circ}$.  This is plotted by the dotted line 
in the lower panel of Figure \ref{fig:ppdot}. Thus, the scatter in data is a 
combination of X-ray production efficiency and viewing angle effects.
The two ``outliers" namely, AXP 1E2259$+$586 and AXP 4U0142$+$615 
 can be explained as those AXPs/SGRs born in the 
propeller regime so that they are surrounded by a degenerate torus instead
of a shell, and are discussed in more detail in paper II. 

\subsection{The two-component blackbody\label{sec:2bbs}}

Our model provides an explanation to the two component blackbody spectrum 
that may be seen in SGRs/AXPs (i.e.~Israel 2002).
This is realized by considering the blackbody temperature as set by the X-ray continuum luminosity, 
$L_{\rm X} = 4\pi R_{\rm QS}^2  T_{\rm BB}^4$, to be\footnote{Equation (\ref{eq:Tbbsgr})
 is a direct measure of the star's radius once $T_{\rm BB}$ (or $L_{\rm X}$)  and $\dot{P}$ are measured.
 This could become crucial for deriving the Mass-Radius relationship for these objects.
  The mass (more precisely $M/R$) could be derived from photon redshifts.}
\begin{equation}
\label{eq:Tbbsgr}
T_{\rm BB}\simeq 0.41\ {\rm  keV}\ \eta_{\rm X}^{1/4} \left(\frac{\dot{P}_{-11}}{R_{\rm QS, 10}}\right)^{1/2}\ ,
\end{equation}
 where we made use of equation~(\ref{eq:lx}) for $L_{\rm X}$ and 
 adopted the fiducial value $R_{\rm QS}=10$ km. 

The effective shell temperature in our model is  also
set by the X-ray continuum luminosity, 
$L_{\rm X}\sin\theta_{\rm B} = 4\pi R_{\rm m}^2 \sin \theta_{\rm B} \sigma T_{\rm sh,eff.}^4$, and is
\begin{equation}
T_{\rm sh,eff.} \simeq 0.41 \ {\rm keV}\ \eta_{\rm X}^{1/4}\left(\frac{\dot{P}_{-11}}{R_{\rm m, 10}}\right)^{1/2}\ .
\end{equation}
Thus there is two temperatures from two different emitting blackbodies, namely the quark star and the
surrounding shell. 

Whether or not the blackbody component from the shell
could actually be observed or not hasn't been considered.  
If it can not, then the magnetic field strengths around
CFL quark stars is still at a magnitude that
can produce the synchrotron break (i.e. blackbody plus power law) 
that is usually assumed for SGRs/AXPs.  
We only assert that if the blackbody emission from the shell could be observed, 
then its temperature is such that it correlates well with a two-component
blackbody spectrum model.

\section{Period changes and torques}\label{sec:periodchanges}

Angular momentum exchange between the shell and the star
occurs during the shell's inward and outward movements,
and would manifest itself as changes in rotational period and period derivative during bursts.
We start with the inward drift where the moment of inertia of the star decreases, 
causing it to spin-up.
  
\subsection{Inward Shell Drift: Spin-up}

In our model the inward shell drift occurs while the vortex (i.e. magnetic flux) expulsion 
mechanism, as discussed in Niebergal et al. (2006), is coupling the spin period and 
field decay as described in equations~(\ref{eq:tau} - \ref{eq:b_evol}).
The spin-up torque from a decreasing moment of inertia occurs slowly during 
the magnetic field decay timescale over time $\alpha \tau$, 
and is negligible compared to the spin-down due to vortex expulsion.

 The shell eventually reaches an inner radius where it is unstable in the magnetic field
geometry, and pieces start to break off above $\theta_{\rm B}$. 
As the pieces fall in along field lines to the polar regions and collide with the star, 
angular momentum is transferred.
This transfer and the change in moment of inertia, $\Delta I$, 
as we show below would correspond to a star spinning-up during bursts (i.e. accretion
of chunks)\footnote{In the cases where the shell moves in and out without
pieces breaking off, there would be associated $\dot{P}$ variations
without SGR bursts.}. 
The change in moment of inertia from the increase in mass and radius of the star is given by,
\begin{equation}
  \frac{\Delta I}{I} = \frac{5}{3}\frac{\Delta m}{M_{\rm QS}}\ ,
\end{equation}
where the accreted matter is converted into CFL matter at the star's density.
 Taking into account the angular momentum, $\Delta L$,  
from the infalling (from $R_{\rm m}$) shell material we get,
\begin{equation}
 \frac{\Delta P}{P} = \frac{\Delta I}{I} - \frac{\Delta L}{L} 
 = \frac{5}{3}\frac{\Delta m_{\rm{sh}}}{M_{\rm QS}}
   \left(1 - \frac{3}{2}\frac{R_{\rm m}^2}{R_{\rm QS}^2}\right) \ ,
\end{equation}
where a quark star mass and radius of $M_{\rm QS}=1.4M_{\odot}$ and $R_{\rm QS}=10~\rm{km}$
is used, and will be assumed for the remainder of this section.
Making use of equations~(\ref{aradius} \& \ref{eq:mloss}) for $R_{\rm m}$ and $\Delta m$ 
respectively one arrives at,
\begin{equation}\label{eq:deltap}
    \frac{\Delta P}{P} \simeq -1.8 \times 10^{-6} f\left(\alpha\right) 
    B_{15}^2
    \left(1 - \frac{3}{2}\frac{R_{\rm QS}^2}{R_{\rm m}^2}\right) \ .
    \label{eq:spinsown}
\end{equation}
Using equation~(\ref{eq:3kppdot}), the equation above can be recast into 
\begin{equation}
  \frac{\Delta P}{P} \simeq - 4.75\times 10^{-7} f(\alpha) P_{10} \dot{P}_{-11}\ ,
\end{equation} 
where $P_{10}$ is the spin period in units of $10~\rm{s}$, $\dot{P}_{-11}$ is the 
period derivative in units of $10^{-11}$, and we had assumed that $R_{\rm m} \gg R_{\rm QS}$.

If the accretion occurs during a time interval, $\Delta t_{\rm acr}$, 
 the spin-up rate, $\dot{P}_{\rm acc.}= \Delta P/\Delta t_{\rm acc.}$, can be estimated by dividing both sides of equation~(\ref{eq:deltap}) by $\Delta t_{\rm acc.}$. 
By noting that $\dot{P}_{\rm old} = B^{2}/(3\kappa P)$ we arrive at 
\begin{equation}
 \frac{\dot{P}_{\rm acc.}}{\dot{P}_{\rm old}} = - 132  f\left(\alpha\right)
 \left(\frac{1~\rm{hr}}{\Delta t_{\rm acc}}\right) P_{10}^2  \ .
\label{eq:pdotnew}
\end{equation}  

\subsection{Outward Shell Rebound: Spin-Down}

Following the accretion events, during the fast shell rebound, 
angular momentum is transfered from the star to the shell to keep it 
co-rotating, resulting in an increase in the spin-down rate of the star. 
The magnetic coupling between the star and the shell implies that the
star will lose angular momentum and spin-down.  As the star moves outward
from  radius $R_{\rm in}$ back to $R_{\rm m}$,
conservation of angular momentum of the star/shell system implies
$L_{\rm QS, in}+L_{\rm shell, in} = L_{\rm QS, m} + L_{\rm shell, m}$. That is
\begin{equation}
 \frac{\Delta P}{P} = +  \frac{5}{2}\frac{m_{\rm sh}}{M_{\rm QS}}
 \left( \frac{R_{\rm m}^2}{R_{\rm QS}^2} - \frac{R_{\rm in}^2}{R_{\rm QS}^2}\right)\ ,
\end{equation}
or by using equations~(\ref{aradius} \& \ref{eq:constmass}) one gets,
\begin{equation}\label{eq:spinup}
 \frac{\Delta P}{P} \simeq + 4.1\times 10^{-7} f\left(\alpha\right) P_{10}\dot{P}_{-11} \ ,
\end{equation}
which as it turns out is insignificant as compared to mass loss from the shell's atmosphere.

\subsection{Shell atmosphere loss: Spin-down}\label{sec:nondegen_atmos}

Observationally, an increase in spin-down has been seen to last for $\sim 18$ days,
as in the case of AXP1E2259, and $\sim 80$ days for SGR1900$+$14.  
In our model this is readily explained by the heating of the degenerate shell by the burst,
and the subsequent release of a portion of it's atmosphere.
Because a thin layer on the shell's surface is in the non-degenerate phase, 
a fraction of it is blown away via the propeller mechanism, 
removing angular momentum from the system over time.  

To determine the amount of atmosphere present on the shell during the quiescent phase, 
we consider the critical density,  $\rho_{\rm sh,nd}$, 
of the shell to ensure degeneracy which is found by setting 
$T_{\rm sh,eff.}= T_{\rm Fermi} \simeq  76.2\ {\rm MeV}  \rho_{\rm sh,nd, 10}^{2/3}$,
where the density is written in units of $10^{10}~\rm{g/cc}$. 
So the gas will be non-degenerate below densities of, 
\begin{equation}
  \rho_{\rm sh,nd} \simeq 3.6\times 10^{3}\ {\rm g/cc}\ \left(\frac{kT_{\rm sh,eff.}}{1\ {\rm keV}}\right)^{3/2}\ .
\end{equation}
The scale height of the non-degenerate, upper layers, of the shell can be estimated
to be 
\begin{equation}
H_{\rm sh,nd} \simeq \frac{k T_{\rm sh,eff.}}{(\mu_{\rm Fe} m_{\rm H} g)}\sim
 0.1 \ {\rm cm}\ R_{\rm m, 10}^2 \left( \frac{T_{\rm sh, eff.}}{1 {\rm keV}}\right) \ ,
\end{equation}
where $g = GM_{\rm QS}/R_{\rm m}^2\sim 1.87\times 10^{14}/R_{\rm m,10}^2\ {\rm cm~s}^{-2}$
 is the effective gravity.
Thus, the mass of the non-degenerate portion of the shell is,
\begin{eqnarray}\label{eq:nondegen_mass}
\nonumber  m_{\rm sh,nd} &\simeq& 4\pi R_{\rm m}^2 \sin\theta_{\rm B} \ H_{\rm sh, nd}  \rho_{\rm sh, nd} \\
 &\sim& 3.6\times 10^{15} \ {\rm gm.} \ R_{\rm m,10}^4 
 \left( \frac{kT_{\rm sh, eff.}}{1 {\rm keV}}\right)^{5/2}\ ,
\end{eqnarray}
  where $\theta_{\rm B}=60^{\circ}$ was used.
  
   Following a bright burst, with luminosity $L_{\rm b,44}$ (in units of $10^{44}~\rm{erg~s}^{-1}$), 
the shell gets reheated to temperatures of the order
$T_{\rm sh}(t) \simeq 50\ {\rm keV} (L_{\rm b, 44}(t)/R_{\rm m,10})^{1/4}$.
The corresponding atmospheric (non-degenerate) mass in terms of the burst energy 
(in units of $10^{44}~\rm{erg}$), is then 
\begin{equation}\label{eq:shellatmosphere}
m_{\rm sh, nd}(t) \simeq 6.32\times 10^{19}\ {\rm gm.}\ R_{\rm m,10}^{27/8}\ L_{\rm B,44}^{5/8}(t)  \ .
\end{equation}  

So the trigger mechanism is that the temperature needs
to be high enough for the atmosphere to leak away from the shell,
most likely in the form of a pressure driven wind.
Once it is gravitationally unbound it is kept in co-rotation by the
magnetic field out to the light cylinder, providing an efficient propeller.
The angular momentum per unit mass lost at the light cylinder is $c^2/\Omega$
which gives a new spin-down rate for the quark star of,
\begin{equation}
 \dot{P}_{-11,{\rm new}} \simeq 2 \dot{m}_{12}  P_{\rm 10}^3\ , 
\end{equation}
where the estimated mass-loss rate, $\dot{m}_{12}$, is in units of $10^{12}\ {\rm g\ s}^{-1}$.

The change in period due to mass lost at the light cylinder is also written as,
\begin{equation}
\frac{\Delta P}{P} = \frac{5}{2}\frac{\Delta m_{\rm sh, nd}}{M_{\rm QS}}
\left(\frac{R_{\rm lc}}{R_{\rm QS}}\right)^2 \ ,
\end{equation}
where $R_{\rm lc}$ is the radius to the light cylinder. Assuming
$\Delta m_{\rm sh,nd}= \zeta m_{\rm sh,nd}$ where $\zeta$ is the portion
of the non-degenerate atmosphere lost to the light cylinder and $m_{\rm sh, nd}$
is the peak of the atmospheric mass during the burst.  We get,
\begin{equation}\label{eq:periodchange}
\frac{\Delta P}{P} = 1.3\times 10^{-4} \ \zeta L_{\rm B,44}^{5/8} P_{10}^2 R_{\rm m,10}^{27/8} \ ,
\end{equation}
where in the equation above we use  the peak burst luminosity 
which we represent by the subscript ``B" as to differentiate it from $L_{\rm b, 44}(t)$ .

Because the shell cools, there is decreasing amounts of matter available to drive the increased
spin-down rate. Thus this new rate is temporary and the timescale for it to decay back to the previous rate
can be estimated by, $t_{\dot{P}} = \Delta m_{\rm sh,nd}/\dot{m}$, or,
\begin{equation}
t_{\dot{P}} = 4 ~\rm{yrs}\times \zeta L_{\rm B,44}^{5/8} R_{\rm m,10}^{27/8} 
P_{10}^3 \dot{P}_{-11,{\rm new}}^{-1} \ .
\end{equation}
One should keep in mind  that in reality
the mass loss rate is decreasing in time.  Determining better estimates for this requires
knowledge of $\zeta$ which  is beyond the scope of this paper.

\section{Application to SGRs and AXPs}\label{sec:applications}

\subsection{SGR 1806$-$20: The boxing day SGR}

The most recent and brightest giant flare came from SGR 1806$-$20 on Dec. 27, 2004. 
This flare lasted about 5 minutes and had a peak luminosity of about 
$2 \times 10^{47}\  {\rm erg~s}^{-1}$.
For a distance to SGR 1806$-$20 of $15$ kpc, 
it is estimated that an (isotropic equivalent) energy release of $2 \times 10^{46}$
erg occured in the spike, and $5 \times 10^{43}$ erg in the tail. 

In our model the energy is readily accounted for. The observed $P$ and $\dot{P}$
imply (from Eq.~\ref{eq:3kppdot}) that $B \sim 1.3\times 10^{15}$ G, 
which translates to an energy of $3.6\times 10^{46}$ erg for $\alpha =1$ 
and $R_{\rm m}= 20$ km (see Eq.\ref{eq:eburst}). 

In the year leading up to the SGR 1806$-$20 giant flare, 
well-sampled X-ray monitoring observations of the source with the Rossi X-ray Timing 
Explorer (RXTE) indicated that it was also entering a very active phase, emitting more 
frequent and intense bursts as well as showing enhanced persistent X-ray emission which was, 
a prelude to the unprecedented giant flare.  
 In our model, this prelude would correspond to increasing amounts of shell 
pieces breaking off above $\theta_{\rm B}$ as the shell moves in rapidly
closer to the star.  This situation is likely only when the quark star's magnetic field
is strong such that it is decaying quickly, implying the object is young, 
which is the case with SGR 1806$-$20.  Furthermore, because $\sim 20\%$ of the shell's
mass was lost in this giant flare, the leftover smaller shell, once re-adjusted
to $R_{\rm m}$ following the event, will have less capability to produce flares of the same magnitude
as the previous one (see eq.(\ref{eq:erecur})).

The sharp initial rise of the main spike
in the boxing day flare was of the order of 1 millisecond (Palmer et al. 2005). 
In our model, a lower limit on the rise time, 
\begin{equation}
  t_{\rm rise,min }  =  \frac{\Delta R_{\rm m}}{v_{\rm ff}}  \ ,
\end{equation}
is determined by the time it takes a typical chunk of radius $\Delta R_{\rm m}$
to cross the star surface, and be converted to CFL quark matter. 
Larger rise times would correspond to groups of shell pieces accreting simultaneously.
To a first approximation we assume that the shell piece
crosses the quark star surface at maximum (free-fall) velocity of, 
$v_{\rm ff}= \sqrt{GM_{\rm QS}/R_{QS}}$. 
Making use of equations~(\ref{eq:drm}, \ref{eq:eburst} \& \ref{eq:3kppdot}) we get,
\begin{equation}\label{eq:ffall}
 t_{\rm rise, g} >  0.5~\mu s \ \frac{\sqrt{P_{10}\dot{P}_{-11}} 
 \left(\eta_{0.1} f\left(\alpha\right)\right)^{1/4}}{E_{\rm b, 47}^{1/4}}\ ,
\end{equation}
 where the burst energy, given in units of $10^{47}\ {\rm erg}$,
 is representative of a giant burst or the case where $\alpha \ge 1$ 
(denoted by subscript ``g" in the equation).
 
For regular bursts ($\alpha \ll 1$)  we find
\begin{equation}
 t_{\rm rise, r} > 0.9~\mu s \ \frac{\sqrt{P_{10}\dot{P}_{-11}}
 \left(\eta_{0.1} \alpha_{-6} \right)^{1/4}}{E_{\rm b, 40}^{1/4}}\ ,
\end{equation}
where $\alpha_{-6}= \alpha/10^{-6}$ was used and the burst energy
is given in units of $10^{40}$ erg. However, in the case of small bursts the total rise time
can be larger, as multiple shell pieces are likely to be falling into the star simultaneously.

\subsection{SGR 1900$+$14}

The rotation period has been studied in detail for SGR 1900$+$14 during bursting
(i.e.~Woods et al. 1999 and Palmer 2002), which allows us to test these features of our model.
The increase in period as given in our model (Eq.~\ref{eq:periodchange}) following the burst is,
$\Delta P/P \sim 3.9\times 10^{-5}\times  \zeta R_{\rm m,10}^{27/8}$ using observed values for energy and period.   Also, observed values for period and the new period derivative following the August 27th burst 
give a mass loss rate of $\dot{m} \sim 5\times 10^{13}~\rm{g/cc}$, and an upper limit for the
recovery time of $t_{\dot{P}}\sim 15\ {\rm  days}\ \times \zeta  R_{\rm m,10}^{27/8}$.

The magnetic equilibrium radius was estimated to be $80~\rm{km}$ due to the presence
of iron line emission (Strohmayer \& Imbrahim 2000) in which case for $\zeta\sim 0.004$ we get 
  $\Delta P/P\sim 1.5\times 10^{-4}$ and $t_{\dot{P}}\sim 67$ days.
 However as discussed in \S~\ref{sec:ironline} for an ionized atmosphere, redshift
 correction could move the radius to $30~\rm{\rm km}$ in which case a $\zeta\sim 0.1$
 gives $\Delta P/P\sim 1.4\times 10^{-4}$ and $t_{\dot{P}}\sim 61$ days.
This is in good agreement with what can be inferred
from observations.

\subsubsection{Iron line emission during bursts}\label{sec:ironline}

Observations of Fe fluorescence lines in SGR 1900$+$14 (Strohmayer \& Ibrahim 2000)
are a key feature in our quark star model.  As discussed in Ouyed et al. (2002) and Ker\"anen,
Ouyed, \& Jaikumar (2005),
Fe and heavier metal production in the ejecta is significant during the formation of a quark star.
Strohmayer \& Ibrahim state that the distance from the star's surface, 
where the emission is created needs to be at least, $h\sim 80~\rm{km}$, 
to account for the lack of redshift\footnote{If the line
is from ionized iron, the rest energy can increase up to
6.7 keV thus a redshift up to 5\% is allowed, reducing the lower
limit on the distance of the iron emitting gas from the star to $\sim 30$ km. }.  
This distance gives an excellent confirmation to our model, 
as it corresponds to the distance of the shell which possess an Fe-rich atmosphere.

The atmosphere's column density can be calculated from equation~(\ref{eq:nondegen_mass}) to be, 
\begin{equation}
 N_{\rm Fe} 
 = 3.8\times 10^{24}\ {\rm cm}^{-2}\ R_{\rm m,10}^2 
 \left( \frac{kT_{\rm s, eff.}}{1 {\rm keV}} \right)^{5/2}\ .
\end{equation}
This is sufficient to produce a strong $K_{\alpha}$ iron
emission line (equivalent width $>$ 100 eV), when illuminated.
Thus in our model, the Fe flourescent line is produced during bursts by the illumination 
of the shell's atmosphere.

\subsection{X-ray to Infrared Flux Ratio}\label{sec:XvsIR}

Another observed feature in both SGRs and AXPs is the X-ray to infrared flux
ratio correlations.
The X-ray flux during the quiescent phase 
(induced by the magnetic field decay) and/or during the
bursting phase  (from accretion events)  can be thermally reprocessed by the
shell into the far infrared.  However, the efficiency is
too low due to the small shell size and cannot account for the observed
values of $(\nu F_{\nu})_{\rm X}/(\nu F_{\nu})_{\rm IR}$ which are $\sim 150$ for
SGRs and $\sim 1500$ for AXPs  (e.g.~Israel et al. 2003).
 
Alternatively one expects a much higher infrared
flux related to vortex annihilation, which is always
present in this model. The mechanism is  synchrotron 
emission from mildly relativistic electrons accelerated by the magnetic
reconnection events induced by vortex annihilation (see simulations in Ouyed
 et al.  2005). The X-rays are from the high energy electrons and the
infrared from the low-energy electrons.
 
For a power-law electron energy distribution $N(E)\propto E^{-\gamma}$
the optically thin synchrotron flux is given by $F_{\nu}\propto \nu^{-(\gamma -1)/2}$
(Longair 1992). This implies
\begin{equation}
 \frac{(\nu F_{\nu})_{\rm X}}{(\nu F_{\nu})_{\rm IR}} =  (\frac{\nu_{\rm X}}{\nu_{\rm IR}})^{(3-\gamma)/2}
 \simeq (2\times 10^4)^{(3-\gamma)/2}\ .
\end{equation}
For a typical index $\gamma=2$ this yields
\begin{equation}
  \frac{(\nu F_{\nu})_{\rm X}}{(\nu F_{\nu})_{\rm IR}} \sim 140 \ .
\end{equation}
In this model both $L_{\rm X}$ and $L_{\rm IR}$
are proportional to $\dot{P}^2$ (see Eq.~\ref{eq:lx}) which also implies  $L_{\rm IR}/L_{\rm X}$ 
is a constant.    Synchrotron self-absorption at lower frequencies, low-energy electron losses, 
or a smaller index $\gamma$, all increase  the $F_{\rm X}/F_{\rm IR}$ ratio, and 
may be the reason why   AXPs show larger ratios than SGRs. 
During bursting episodes the vortex annihilation rate
increases due to the spin-up induced by accretion increasing both $L_{\rm X}$
and $L_{\rm IR}$, thus they remain correlated during bursts.

\subsection{The hard spectrum/component in our model\label{sec:hard}}

Three of the four known SGRs have had hard spectrum (with photons
 in the MeV energy) giant flares.  Before showing how this can be accounted
 for in our model, recall that in the case of quark
 stars the surface emissivity of photons with energies below
$\hbar\omega_{\rm p}$~$\simeq$23~MeV ($\omega_{\rm p}$: 
electromagnetic plasma frequency) is strongly suppressed 
(Alcock, Farhi, \& Olinto 1986; Chmaj, Haensel, \& Slomi\'nsli 1991; 
Usov 1997). In Vogt, Rapp, \& Ouyed (2004) it was shown that 
 average photon energies in quark stars in the CFL phase 
at temperatures $T$ are $\sim$3$T$. Therefore, as soon
as the surface temperature of the star cools
below $T_{\rm a} = \hbar\omega_{\rm p}/3\simeq 7.7$~MeV, 
the photon emissivity is highly attenuated.
 This is studied in more details in Ouyed, Rapp, \& Vogt (2005)
  where it was demonstrated that for temperatures above 7.7
  MeV, neutrino cooling is dwarfed by the photons;
i.e., photon emission/cooling dominates as long as the star cools from 
its initial temperature $T_0>7.7$ MeV to $T_{\rm a}=7.7$ MeV. For temperatures
below 7.7 MeV, cooling is dictated by the slower neutrino processes.
 
 To a first approximation, the increase in the star's temperature, $\Delta T_{\rm QS}$, 
  following the accretion of shell pieces can be written as,
 \begin{equation}
 4\pi R_{\rm QS}^{2}\Delta R (1-\sin\theta_{\rm B}) c_{\rm v} \Delta T_{\rm QS} \simeq
  \eta \Delta m_{\rm sh} c^2 \ ,
 \end{equation}
 where $c_{\rm v}= 7.8\times 10^{16} T_{\rm MeV}^3\ {\rm erg\ cm}^{-3}\ {\rm K}^{-1}$ is the star's
 specific heat as given in Ouyed, Rapp, \& Vogt (2005). Also,
  $\Delta R$ is the thickness\footnote{The heat penetration depth is
  given as $\Delta R = c_{\rm s} t_{\rm ff}$ where $c_{\rm s}\sim c/\sqrt{3}$
  is the CFL sound speed and $t_{\rm ff}$ is the free fall timescale given by equation
  (\ref{eq:ffall}). We get $\Delta R \simeq 10^{4}\ {\rm cm}\ B_{15}\times (\eta_{0.1}f(\alpha)/E_{\rm b, 47})^{1/4}$.}  of the heated region in the polar
   caps. Since the pieces fall
 off above the neutral line (i.e. $\theta > \theta_{\rm B}$), the polar regions are the ones that are
 primarily heated in the process. The corresponding increase in temperature
  (assuming the initial star temperature is in the keV range) is then,
 \begin{equation}
 T_{\rm QS} \simeq  10~{\rm  MeV} \frac{E_{\rm b, 45}^{3/16}}{\left((1-\sin\theta_{\rm B}) \eta_{0.1}^{1/4} f^{1/4}(\alpha) R_{\rm QS, 10}^2  B_{15} \right)^{1/4}}\ .
 \end{equation}
 For $\theta_{\rm B}= 60^o$, $\alpha =1$, and $B_{15}\sim 1$,
 this implies that only accretion events with energy exceeding 
$E_{\rm b, c}\sim 2.8\times 10^{43} $ erg are capable
  of reheating the star above 7.7 MeV, allowing MeV photons to be generated and escape
  the star.  Thus our model naturally explains why only the rare giant bursts emit a hard-spectrum.

   The escaping photons are thermalized and cool the star at a rate
   given by $4\pi (1-\sin\theta_{\rm B}) R_{\rm QS}^2 \Delta R c_{\rm v} \partial T_{\rm QS}/\partial t
   = - 4\pi  R_{\rm QS}^2 \sigma T_{\rm QS}^4$ which leads to a cooling time of,
\begin{equation}
 t_{\rm c, QS} \sim 0.03\ {\rm ms}~ 
 \frac{(1-\sin\theta_{\rm B})\eta_{0.1}^{1/4}f^{1/4}(\alpha)B_{15}}{E_{\rm b,45}^{1/4}} 
 \log{\frac{T_{\rm QS}}{T_{\rm a}}}\ .
\end{equation}
   As the high energy photons escape into the magnetosphere
   they interact with lower energy photons and $(e^+e^-)$ pairs, and so become
   thermalized to about $\sim 1$ MeV.
 The $(e^+e^-)$ pairs could be due to 
the radiation from  the magnetic field decay which occurs 
outside the surface of the quark star. We note that the radiation produced by accretion
is from the conversion of baryons to quarks at the surface which also
releases photons at the surface (half of this radiation heats the quark 
matter inside the star while the other half is radiated promptly). 
We have not studied these  
emission mechanisms  in details but 
 we are currently investigating mechanisms for thermalizing the escaping radiation,
 and the nature of the resulting spectrum. We  can simply argue for now that 
 the radiation mechanisms should be  similar to those for a pair-dominated  fireball
  in an optically thick environment.

\subsection{Summary of outstanding SGR/AXP questions}\label{sec:summary}

We summarize this section by attempting to test our model against the open issues
 related to SGRs/AXPs as discussed in the literature (i.e.~Israel 2006). These open issues
 are enumerated below. 

\begin{enumerate}

\item Are the SGRs/AXPs engines born with milliseconds periods?

Allen \& Hovarth (2004) and Vink \& Kuiper (2006) give good evidence in two cases for
normal energy supernova shells around an AXP (1E1841$-$045) and an SGR (0526$-$66), 
which implies birth periods larger than tens of milliseconds. In our model
there is no need for rapid rotation at birth.

\item What differentiates the two types (giant and regular) of bursts?
 How does the boxing day, SGR 1806$-$20 event fit in this picture?

 A giant burst is due to the shell losing a larger
fraction of its mass as it moves towards  the star,  parameterized in our model by $\alpha\sim 1$. 
Regular bursts, $\alpha \ll 1$, are due to smaller pieces breaking off the neutral line as the shell oscillates
around its equilibrium position.  For SGR 1806$-$20 we argue the boxing
day event  is one of the
first events experienced by this object following its birth. 
Moreover, our model predicts (via eqn.~\ref{eq:erecur}) that 
we should continue to see a decreasing trend in both the burst intensity 
and frequency going from younger to older objects.

\item Why are the IR and X-ray variability correlated during flares? 
Is the far-IR emission due to a passive disk? Why passive? Why disk?

The far-IR emission and the correlation with the X-ray
emission can simply be explained as synchrotron emission
from the high-energy and low-energy electrons  (see \S~\ref{sec:XvsIR}). 
Our disk is in fact an iron-rich shell.  It is passive because it is degenerate 
and for most of its lifetime it remains in equilibrium at $R_{\rm m}$, co-rotating with the star.
 
\item Is there any connection between AXPs/SGRs and high-B radio pulsars?

AXPs/SGRs are quark stars and high-B radio pulsars are neutron stars
that have not gone through a QN phase, probably because their core
densities never reached deconfinement values (see Staff, Ouyed, \& Jaikumar 2006).

\item What is the origin of the hard X-ray spectrum in AXPs/SGRs?

The hard X-ray spectrum in our model can be explained as MeV photons
generated in the outer layers of the star. These photon bursts can only  
occur for accretion events capable of heating the star above 7.7 MeV. 
We find that only bursts with energies above $10^{43}$ erg can do so 
(see \S~\ref{sec:hard}), thus explaining why only the giant SGR flares show a hard spectrum.

\item Why do the spin-down ages and the supernova ages differ?

In our model the difference can be explained by the time
it takes the neutron star to reach deconfinement/nucleation densities as discussed in 
\S~\ref{sec:ageestimate}.  Simply put, the supernova age is the time for the neutron
star to reach quark deconfinment densities and experience a quark-nova, plus the time
needed for strange quark nucleation ($t_{\rm SN} = t_{\rm QN} + t_{\rm nucl.}$).

\item What are the progenitors of AXPs/SGRs and why are they located in dense ISM?

Massive stars (near the black hole line) exploding in high density
ISM are most likely to lead to massive compact remnants. The 
high density ISM will confine the massive progenitor
winds much closer to the star, causing the deceleration of the blast wave, and initiating
the reverse shock inside the remnant (Truelove \& McKee 1999).
This would lead to more massive compact stars which are more likely to turn
directly into quark stars.

\end{enumerate}

\section{XDINs in our model}\label{sec:xdins}
 
In our model X-ray Dim Isolated Neutron Stars (XDINs) are old SGRs/AXPs 
that have gone through their most active bursting
phase and are left with a thin shell in stable equilibrium
at $R_{\rm m}$. We start by summarizing the
observed and measured features of XDINs before we apply our model
to these intriguing objects.

\subsection{Properties of XDINs}

These dim ($L_{\rm X}\sim 10^{31}\ {\rm erg\ s}^{-1}$) isolated neutron stars are nearby at around  100- 300 pc and show no SNR association. Three of them have known proper motions that are too fast to accrete. The main common properties of the ``magnificent seven" can be summarized as follows (see Haberl 2005 and references therein): (i)  {\it The blackbody}:  The X-ray spectra of XDINs obtained by the ROSAT PSPC are all
 consistent with blackbody emission. The  
  soft  X-ray spectra have  a temperature in the range of 40-110 eV.
 They show no non-thermal component; (ii) {\it The Absorption lines}:
The XMM-Newton spectra can be best modeled with a Planckian continuum including a broad, Gaussian shaped absorption line. The line centroid energies are in
 the range 100-700 eV. The depth of the absorption line (or the equivalent width)
 was found to vary with  pulse phase. It has been suggested in the literature that these absorption 
 lines can be best explained as proton cyclotron resonance absorption features in
 the 0.1-1 keV band  with field strength in the range of $2\times 10^{13}$-$2\times 10^{14}$
 G (Zane et al.2001; Zavlin\&Pavlov 2002) with the line broadening explained as 
 due to the variation of the magnetic field over the neutron star surface;  (iii) {\it The optical excess}:
They show optical excess compared to the X-ray blackbody. In other words at optical
wavelengths they show a factor of about 3$-$14 excess when compared to the
extrapolation from X-rays (Pons et al. 2002; Motch\&Haberl 1998; 
 Haberl et al. 2004; van Kerkwijk et al. 2004); (iv) {\it The Lack of radio-emission}:
It has been argued that the lack of pulsed radio-emission  is because
their radio beam is very narrow due to the large light cylinder radius (i.e.,
large periods). However, there exist radio-pulsars with similar magnetic
field strengths and periods (e.g. Camilo et al. 2000) that are active in the radio.
In our model, the lack of radio pulsation is due to the fact that quark
stars in the CFL phase, unlike neutron stars,  become  aligned-rotators 
following the QN (see Ouyed et al. 2005).  We now go on to discuss the remaining properties.

\subsection{The two-component blackbody}

As discussed in \S \ref{sec:2bbs}, 
 the first blackbody temperature is 
set by the X-ray continuum luminosity from the surface of the quark star, 
\begin{equation}
\label{eq:Tbb}
T_{\rm BB} \simeq  41.0 \ {\rm  eV}\ \left(\frac{\dot{P}_{-13}}{R_{\rm QS, 10}}\right)^{1/2}\ .
\end{equation}
 Note that  the spin-down is in units
 of $10^{-13}$ s/s  reflective of what has been measured for  XDINs.
  The observer would see an  emitting region with a corresponding blackbody radius
  $R_{\rm BB} = R_{\rm QS} \sqrt{1-\sin\theta_{\rm B}}\sim 4.4$ km using our fiducial
  value $\theta_{\rm B}=60^o$. Interestingly, these are the same values as inferred for
   RX J1856 (e.g. Burwitz et al. 2003) and RXJ0720 (e.g. Haberl 2004).
   
  Similarly, the effective shell temperature (the second blackbody)  is 
\begin{equation}
\label{eq:Teff}
T_{\rm sh,eff.} \simeq  41.0 \ {\rm  eV}\ \left(\frac{\dot{P}_{-13}}{R_{\rm m, 10}}\right)^{1/2}
 = T_{\rm BB} \left( \frac{R_{\rm QS, 10}}{R_{\rm m, 10}} \right)^{1/2}\\ .
\end{equation} 

\subsection{Absorption lines}

Since $R_{\rm m} > R_{\rm QS}$ in our model it implies  $T_{\rm sh, eff.} < T_{\rm BB}$
which means that the shell will act as an absorber of the hotter X-ray blackbody.
 The maximum density, $\rho_{\rm s,nd}$, of the shell below
  which the gas is non-degenerate is found by setting $T_{\rm sh,eff.}= T_{\rm Fermi}$, or, 
 $ \rho_{\rm sh,nd} \simeq 54\ {\rm g\ cm}^{-3}\times \left(T_{\rm sh,eff.}/100\ {\rm eV}\right)^{3/2}$.
  The scale height of the non-degenerate, upper layers, of the shell can be estimated to be
  \begin{equation}
   H_{\rm sh,nd} \sim\frac{ k T_{\rm sh,eff.}}{\mu_{\rm Fe} m_{\rm H} g_{\rm s}}\sim
   0.01 \ {\rm cm}\ R_{\rm m, 10}^2 \left( \frac{T_{\rm sh, eff.}}{100 {\rm eV}}\right)\ ,
   \end{equation}
  implying a  mass of the non-degenerate portion of the shell of 
  \begin{eqnarray}
  m_{\rm sh,nd} &=& 4\pi R_{\rm m}^2 \sin\theta_{\rm B} \ H_{\rm sh, nd} \rho_{\rm s, nd}\\\nonumber
  &\sim& 3.1\times 10^{12} \ {\rm g} \ R_{\rm m,10}^4 \left( \frac{T_{\rm sh, eff.}}{100 {\rm eV}}\right)^{5/2}\ ,
  \end{eqnarray}
   with an optically thick column density,
 \begin{eqnarray}
 \label{eq:NFe}
  N_{\rm Fe} &\simeq& \frac{H_{\rm s,nd}\rho_{\rm s,nd}}{(56 m_{\rm H})} \sim 2.1\times 10^{18}\ {\rm cm}^{-2}\ R_{\rm m,10}^2 \left( \frac{T_{\rm sh, eff.}}{100 {\rm eV}} \right)^{5/2}\\\nonumber 
   &\sim& 2.1\times 10^{18}\ {\rm cm}^{-2}\ R_{\rm m,10}^{3/4} \left( \frac{T_{\rm BB}}{100 {\rm eV}} \right)^{5/2}\ .
\end{eqnarray}
  Iron photospheric models have been calculated by Rajagopal
et al. (1997) for  magnetic field strengths as derived for XDINSs in
our model (i.e.
 $B=\sqrt{3\kappa P\dot{P}}\sim 10^{13}$ G). Despite the fact that these
 calculations are only done in Hartree-Fock approximation
the results  show absorption features in the blackbody spectra, at energies  300 eV and above,
 reminiscent of the lines in XDINSs (see Figure 3 in Rajagopal  et al. 1997;
see also discussion in Neuhauser et al. 1987). Careful attention to Figure 3 in Rajagopal et al. (1997)
 shows that the lines start to disappear at temperature between $30$ eV and $80$ eV.
   Interestingly, this temperature effect may  explain the absence of 
    absorption line in the two XDINSs showing the lowest blackbody temperature
     namely, RX J0420.0$-$5022 with
     $kT_{\rm bb}\sim 40$ eV and RX J1856.5$-$3754 (the brightest or closest
     of the XDINs) with $kT_{\rm bb}\sim 60$ eV.  We thus suggest  that the
iron shell in our model provids the conditions necessary to explain  
the absorption lines as iron lines. In this respect, the iron atmosphere explanation
 of the spectrum suggests that  younger (i.e. hotter) objects of the same type should be observed with a richer
 line spectrum.

 \subsection{The optical excess}

The shell's solid angle, $4\pi\sin\theta_{\rm B}$, is
 essentially   providing the excess optical emission. 
  In the simplest case, one assumes a scattering atmosphere isotropizes
  the optical emitted flux from the QS (i.e. blackbody tail) and the shell. Then
 the ratio between the total  optical flux  and the optical contribution from the tail
 of the blackbody can be expressed as
  \begin{eqnarray}
   \frac{F_{\rm opt., tot.}}{F_{\rm opt., tail}}  &=& 1 + \frac{F_{\rm opt.,sh}}{F_{\rm opt., tail}}\\\nonumber 
   &=& 1 + \frac{4\pi \sin\theta_{\rm B}}{4\pi (1-\sin\theta_{\rm B})}
   =\frac{1}{1-\sin\theta_{\rm B}}\sim 7\ .
   \end{eqnarray}
    In other words, the  optical excess is a direct
   measure of the angle subtended by the shell. This also means
    a one-to-one correspondence between the
   optical excess and the size of the emitting spot in our model  expressed as
 \begin{equation}
 \frac{F_{\rm opt., tot.}}{F_{\rm opt., tail}} = \frac{1}{1-\sin\theta_{\rm B}} = \left(\frac{R_{\rm QS}}{R_{\rm BB}}\right)^2\ ,
\end{equation}
 For example for RXJ1856  $R_{\rm BB}\sim 4.4$ km has been derived from observations. Assuming
 a quark star radius\footnote{It was suggested that the minimum radius of RXJ1856 might exceed 14 km thus favoring stiff equations of state (Tr\"umper 2005). We argue that the inferred radius is in fact  the location of the iron shell. Indeed, the temperature of the cool component was measured to be $< 33$ eV at the $3\sigma$
level (Burwitz et al. 2003). In our model, it implies  $T_{\rm sh,eff.} < 33$ eV or   $R_{\rm m} >  (T_{\rm BB}/T_{\rm sh, eff.})^{2} R_{\rm QS}\simeq (60./33.)^{2} R_{\rm QS}$. That is,  $R_{\rm m} > 40 $ km assuming $R_{\rm QS}\sim 10$ km.} of about 10 km, our model would then predict an optical excess of $ (10/4.4)^2\sim 5$ which is  very close to the factor 5-7 measured from the current optical data and the LETG spectrum (Haberl 2004). The radius of the emitting area in the case of RXJ0720.4 has also been estimated to be
 $4.4$-$4.8$ km (for a distance of 300 pc)  which should give an optical excess of $\sim 5$ in
  our model and similar to  what is observed (see Table \ref{tab:xdins_table}).

\section{Conclusion}\label{sec:conclusion}

In this paper we present a new model for SGRs and AXPs
with possible applications to XDINs. This novel idea relies
on the formation of bare CFL quark stars (within the quark-nova
scenario) as the underlying engine, with the parent neutron
star's crust material surrounding it. Despite the 
simplifications we have employed to study the dynamics of this ejected
shell, specifically how it reaches the equilibrium radius and
how it evolves once it is there, our model has a number
of attractive features that can account for many observed
SGR, AXP, and perhaps XDIN properties.

Missing from our study is the proper treatment of
three-dimensional instabilities acting on the shell,
such as Raleigh-Taylor or oscillations.  Due to the
high conductivity of the shell we expect the
Raliegh-Taylor to not operate on very fast
timescales, and any radial oscillations to be severly damped
as the conducting shell will maintain the magnetic flux
enclosed within.  However, azimuthal oscillations
may have a significant effect.
For a more accuarate description, the dynamics of the shell will require
numerical simulations, and is left as future work.

\begin{acknowledgements}
We thank  K. Mori and P. Jaikumar  for insightful discussions. 
This research is supported by grants from the Natural Science and
Engineering Research Council of Canada (NSERC).
\end{acknowledgements}

\begin{appendix}

\clearpage

\setlength{\voffset}{2cm}

\onecolumn
\begin{landscape}
\begin{longtable}{cccccccccccc}
\caption{Observed features of SGRs and AXPs (Only AXPs and SGRs with measured $P$ and $\dot{P}$ are listed)$^{+}$
\label{Observations}} \\
\hline\hline
name & age$_{\rm SNR}$ (kyr) & P (s) & $\dot{P}_{-11}$ (s/s) & $B_{p\dot{p}}^{\ddag}$ (G) & 
$L_x$ (erg s$^{-1}$) & $ E_{\rm g, burst}$ (erg) & $t_{\rm g, rise}$ (ms) & 
$E_{\rm n, burst}$ (erg) & $t_{\rm n, rise}$ (ms) & $\Delta P/P$  \\
\hline
\endhead
\hline
\multicolumn{11}{l}{
$^{+}$ From the SGR/AXP online catalogue at :
 http://www.physics.mcgill.ca/$^{\sim}$pulsar/magnetar/main.html} \\
\multicolumn{11}{l}{$^{++}$ Objects are listed in order of increasing estimated age (using our model; see \S~\ref{sec:ageestimate});} \\
\multicolumn{11}{l}{$^{\ast}$ transient AXPs;} \\
\multicolumn{11}{l}{$^{\ast\ast}$ Increase in spin-down rate with $\dot{P}_{\rm new} \sim 2.3\dot{P}_{\rm old}$ for about 80 days;} \\
\multicolumn{11}{l}{$^{\dag}$ Very sudden spin-up during the burst with $\dot{P}_{\rm new} \sim 2 \dot{P}_{\rm old}$ for about  18 days;} \\
\multicolumn{11}{l}{$^{\ddag}$ Derived in our model from $B=\sqrt{3\kappa P\dot{P}}$;} \\
\multicolumn{11}{l}{$^{\dag\dag}$ 80 bursts with energies ranging from $3\times 10^{34}$ to $5\times 10^{36}$.} \\
\hline
\endfoot

 sgr1806$-$20     & $17\pm 13$ & 7.49  & 8.3 &  1.28e15  & $3.6\times 10^{36}$ & $ \sim 3.7\times 10^{46}$  & $< 1$ & $10^{36}$-$10^{41}$  & -- & --\\\hline
  sgr1900$+$14    & $20\pm 10$ & 5.17  & 6.1 &   9.12e14 & $2\times 10^{36}$ & $\ge 1.2\times 10^{44}$ & $< 4$ &  $10^{36}$-$10^{41}$& -- & $+ 10^{-4}$$^{\ast\ast}$\\\hline
 sgr0525$-$66     & $10\pm 5$ & 8.05  & 6.5  &  1.18e15 & $8\times 10^{35}$ & $5.2\times 10^{44}$ & $<2$  & $10^{36}$-$10^{41}$ & -- & --\\\hline\hline
 axp1E1048$-$5937 & $20\pm 10$ & 6.45  & 3.81  &   8.05e14 & $3.4\times10^{34}$ & -- & -- & $2.7\times 10^{40}$/$2.8\times 10^{41}$& $21/5.9$ & --\\\hline
 $^{\ast}$taxpxteJ1809$-$1943 & -- & 5.54  &  2.06 & 5.49e14 & $1.6\times10^{36}$ & -- & -- & -- & -- & --\\\hline
 axp1E1841$-$045  & $1.5\pm 1$ & 11.77 & 4.13 &  1.13e15 & $2.3\times10^{35}$ & -- & -- & -- & -- & --\\\hline
 $^{\ast}$taxpJ0100$-$7211& --   & 8.02 & 1.88  &  6.31e14 & $1.5\times10^{35}$ & -- & -- & -- & -- & --\\\hline
 axpRXS1709$-$4009  & $17\pm 13$ & 11.00 & 1.94  & 7.51e14 & $6.8\times10^{35}$ & -- &  -- & -- & -- & --\\\hline
 axp4U0142$+$615  & no SNR & 8.69 & 0.196 &  2.12e14 & $3.3\times10^{34}$ & -- &  -- & -- & -- & --\\\hline
 axp1E2259$+$586  & $10\pm 7$ & 6.98 & 0.0484 &  9.44e13 & $3.8\times10^{34}$ & -- & --& $[3.4\times 10^{37}]^{\dag\dag}$ & (1-100) & $-4\times 10^{-6}$$^{\dag}$ \\ \hline

\end{longtable}
\end{landscape}
\clearpage

\onecolumn
\begin{landscape}
\begin{longtable}{cccccccccc}
\caption{Observed features of the X-ray dim radio-quiet isolated neutron stars$^{\ddag\ddag}$\label{tab:xdins_table}} \\
\hline\hline
ID & period (s) & $\dot{P}_{-13}$ (s/s) & $B_{p\dot{p}}$ & $L_{x}$ (erg s$^{-1}$) &
$kT_{\rm BB}$ (eV) & $E_{\rm EW}$ (eV) & $E_{\rm cen.}$ (eV) & 
Optical Excess &  $R_{\rm bb}$ (km) \\
\hline
\endhead
\hline
\multicolumn{10}{l}{$^{\ddag\ddag}$ See Haberl (2005) and references therein.}
\endfoot
\hline
  RX J0420.0-5022            &  3.45   & $<92$  & -- & $2.7\times 10^{30}$ & 44  & --  &   --  &   $<12$   &   -- \\ \hline
  RX J0720.4-3125            &    8.39  & 0.69  & 3.93e13& $2.6\times 10^{31}$ &  85-95 & 40  &   280  &    $5$    &       4.4-4.8    \\ \hline
  RX J0806.4-4123             &   11.37  & $<18$  & -- & $5.7\times 10^{30}$ & 96  & --  &  306-430   &     --   &        --    \\ \hline
 RBS 1223                            &  10.31    &  $1.12$  & 5.52e13& $5.1\times 10^{30}$ &  86 &  150 &  230-300   &    $<5$   &  -- \\ \hline
 RX J1605.3$+$3249         &    --    &  --  & -- & -- & 96 &  120  &  450-480   &  $<14$     &   --\\ \hline
 RX J1856.5$-$3754          &    --   &  --  & -- & -- & 60  &  --  &  --   &   5-7    &    4.4 \\ \hline
 RBS 1774                             &    9.44   & --  & -- & -- & 101  & --  & $\sim  700$   &   --     &   --\\ \hline
\end{longtable}
\end{landscape}

\end{appendix}

\end{document}